\documentclass[aps,pra,floatfix,showpacs,10pt,twocolumn]{revtex4}
\usepackage{amsfonts}
\usepackage{bbm}
\usepackage{amssymb}
\usepackage[dvips]{graphicx}
\usepackage{amsmath,amsfonts,amssymb,graphics,graphicx,epsfig,color,times,bbm}
\begin{document}

\title{Various Correlations in Anisotropic Heisenberg
XYZ Model with Dzyaloshinskii-Moriya Interaction}

\author{Mamtimin Tursun}
\affiliation{School of Physics and Electronic Engineering, Xinjiang Normal
University, Urumchi 830054, China}
\author{Ahmad Abliz}
\email{aahmad@126.com}%
\affiliation{School of Physics and Electronic Engineering, Xinjiang Normal
University, Urumchi 830054, China}
\author{Rabigul  Mamtimin}
\affiliation{School of Physics and Electronic Engineering, Xinjiang Normal
University, Urumchi 830054, China}
\author{Ablimit Abliz}
\affiliation{School of Physics and Electronic Engineering, Xinjiang Normal
University, Urumchi 830054, China}
\author{Qiao Pan-Pan}
\affiliation{School of Physics and Electronic Engineering, Xinjiang Normal
University, Urumchi 830054, China}

\begin{abstract}
$\bf Abstract$ Various thermal correlations as well as the effect of intrinsic decoherence on the correlations are studied in a two-qubit Heisenberg  XYZ spin chain with the Dzyaloshinski-Moriya ($DM$) interaction along the $z$ direction ($D_z$). It is found that tunable parameter $D_z$ may play a constructive role to the concurrence ($C$), classical correlation ($CC$) and quantum discord ($QD$) in thermal equilibrium while it plays a destructive role to the $C$, $CC$ and $QD$ in the intrinsic decoherence case.
\end{abstract}
\date{\today}

\maketitle

\section{introduction}
Entanglement is a kind of quantum nonlocal correlation and has been deeply studied in the past years \cite{J1,J2,J3,J4}. The quantum discord ($QD$) which measures a more general type of quantum correlation, is found to have nonzero values even for separable mixed states \cite{J5}. $QD$ is built on the fact that two classical equivalent ways of defining the mutual information turn out to be inequivalent in the quantum domain. In addition, $QD$ is responsible for the quantum computational efficiency of deterministic quantum computation with one pure qubit \cite{J6,J7,J8} albeit in the absence of entanglement.

In recent years, the $QD$ has been intensively investigated in the literature both theoretically \cite{J9,J10,J11,J12,J13,J14,J15,J16,J17,J18,J19,J20,J21,J22,J23,J24,J25,J26,J27,J28,J29,J30} and experimentally \cite{J7,J31}. Generally, it is somewhat difficult to calculate $QD$ and the analytical solutions can hardly be obtained except for some particular cases, such as the so-called $X$ states \cite{J10}. Some researches show that $QD$, concurrence ($C$) and classical correlation ($CC$) are respectively independent measures of correlations with no simple relative ordering and $QD$ is more practical than entanglement \cite{J7}. B.Dakic et al \cite{J24} have introduced an easily analytically computable quantity, geometric measure of discord ($GMD$), and given a necessary and sufficient condition for the existence of nonzero $QD$ for any dimensional bipartite states. Moreover, the dynamical behavior of $QD$ in terms of decoherence \cite{J27,J32,J33} in both Markovian \cite{J11} and Non-Markovian \cite{J12,J34,J35} cases is also discussed.

In the previous studies, the QD of a two-qubit one-dimensional XYZ Heisenberg chain with an external magnetic field in thermal equilibrium has been studied,\cite{J36} where many unexpected ways different from the thermal entanglement have been shown. In Ref.\cite{J37} the authors investigated the effect of Dzyaloshinski--Moriya (DM) interaction,\cite{J38} which arises from spin-orbit coupling, on QD in an anisotropic XXZ model and shown that with the increase of the DM interaction the QD gradually reduces at finite temperature. The effect of DM interaction on QD in Heisenberg XY model has also been discussed in Ref.\cite{J16}, in which the authors showed that QD can describe more information about quantum correlation than quantum entanglement. There are interesting papers discussing the QD qualitatively and quantitatively in Heisenberg spin chain models with various factors such as temperature, anisotropies and magnetic field.\cite{J17} In this Letter, we study the QD, CC and $C$ in an anisotropic Heisenberg XYZ model with the DM interaction both in the thermal equilibrium case and the intrinsic decoherence case, and discuss how the DM interaction influence the correlations in such a system. The present study of the correlations in Heisenberg spin chain model will help us to understand the effect of DM interaction on the correlations and the phase decoherence resistance of the correlations more comprehensively.

\section{The Model and Definitions of The Various Correlations}
We consider the anisotropic $XYZ$ Heisenberg model with the anisotropic, antisymmetric $DM$ interaction along the $z$ direction $D_z(\sigma^x_1 \sigma^y_2 \times \sigma^y_1 \sigma^x_2).$ Then the Hamiltonian of such a model can be expressed as
\begin{eqnarray*}
\label{eq:1}\hspace*{9mm}H=\frac{1}{2}[J_{x}
\sigma^{x}_{1}\sigma^{x}_{2}+J_y\sigma_{1}^{y}\sigma^{y}_{2}+J_{z}\sigma_{1}^{z}\sigma_{2}^{z}\qquad
\end{eqnarray*}
\vspace*{-7mm}
\begin{eqnarray}
\qquad\qquad\hspace*{-30mm}+D_z\left(\sigma^x_1 \sigma^y_2 - \sigma^y_1 \sigma^x_2\right)] ,
\end{eqnarray}
where $J_x$, $J_y$ and $J_z$ are the coupling constants; $\sigma^x_i$, $\sigma^y_i$ and $\sigma^z_i$ are the Pauli operators acting on qubit $i ( i= 1, 2).$ In the standard basis ${|\uparrow\uparrow\rangle ,|\uparrow\downarrow\rangle,|\downarrow\uparrow\rangle,|\downarrow\downarrow\rangle} ,$ the Hamiltonian can be expressed in the following matrix form
\begin{eqnarray}
H=\frac{1}{2}\left(
\begin{array}{cccc}
J_z&0&0&J_x-J_y\\
0&-J_z&\beta&0\\
0&\beta^\dag&-J_z&0\\
J_x-J_y&0&0&J_z
\end{array}\right) ,
\end{eqnarray}
where $\beta=J_x+J_y+2iD_z.$
We give a brief overview of various correlation measures. Given a bipartite quantum state $\rho_{AB}$ in a composite Hilbert space $\mathcal {H}=\mathcal {H}_{A}\otimes\mathcal {H}_{B}$, the concurrence \cite{J2} as an indicator for entanglement between the two-qubits is
\begin{eqnarray}
C(\rho_{AB})=\max\{\lambda_1-\lambda_2-\lambda_3-\lambda_4,0\},
\end{eqnarray}
where $\lambda_i(i=1,2,3,4)$ are the square roots of the eigenvalues
of the ``spin-flipped" density operator $R=\rho\widetilde\rho=\rho(\sigma^y_1\otimes\sigma^y_2)\rho^*(\sigma^y_1\otimes\sigma^y_2)$ in descending order. $\sigma_y$ is the Pauli matrix and $\rho^\ast$ denotes the complex conjugation of the matrix $\rho $ in the standard basis ${|\uparrow\uparrow\rangle ,|\uparrow\downarrow\rangle,|\downarrow\uparrow\rangle,|\downarrow\downarrow\rangle}$.
Let us now recalling the original definition of $QD$. In the classical information theory, the total correlations in a bipartite quantum system $(A)$ and $(B)$ are measured by the quantum mutual information defined as
\begin{eqnarray}
{I}(\rho_{A};\rho_{B})=S(\rho_{A})+S(\rho_{B})-S(\rho_{AB}) ,
\end{eqnarray}
where $\rho_{A(B)}=Tr_{B(A)}(\rho_{AB})$ is the reduced density matrix of the subsystem $A(B)$ by tracing out the subsystem $B(A)$. The quantum generalization of the conditional entropy is not the simply replacement of Shannon entropy with Von Neumann entropy, but through the process of projective measurement on the subsystem $B$ by a set of complete projectors $B_k$, with the outcomes labeled by $k$, then the conditional density matrix $\rho_k$ becomes
\begin{eqnarray}
\rho_{k}=\frac{1}{p_k}(\mathbbm{l}_A \otimes {B_k})\rho(\mathbbm{l}_A \otimes {B_k}) ,
\end{eqnarray}
which is the locally post-measurement state of the subsystem
$B$ after obtaining the outcome $k$ on the subsystem
$A$ with the probability
\begin{eqnarray}
p_k=\mathrm{Tr}[(\mathbbm{l}_A\otimes{B_k})\rho(\mathbbm{l}_A\otimes{B_k})] ,
\end{eqnarray}
where $\mathbbm{l}_A$ is the identity operator on the subsystem
$A$. The projectors ${B_k}$ can be parameterized as
${B_k}=V|k\rangle\langle k|V^{\dagger},k=0,1$ and the transform
matrix $V\in U(2)$ \cite{J9} is
\begin{equation}
V=\left(
\begin{array}{cccc}
\cos\theta &   e^{-i\phi} \sin\theta \\
e^{i\phi} \sin\theta & -\cos\theta  \\
\end{array}
\right) .
\end{equation}
Then the conditional Von Neumann entropy (quantum conditional
entropy) and quantum extension of the mutual information can be
defined as \cite{J5}
\begin{eqnarray}
S(\rho|\{B_k\})=\sum_{k} p_k S(\rho_k),
\end{eqnarray}
following the definition of the $CC$ in Ref.\cite{J5}
\begin{eqnarray}
CC(\rho_{AB})=\sup_{\{B_k\}}\{S(\rho_{A}(t))-S(\rho_{AB}(t)|\{B_k\})\} ,
\end{eqnarray}
then $QD$ defined by the difference between the quantum mutual information
$I(\rho_{AB})$ and the $CC(\rho_{AB})$ is given by $QD(\rho_{AB})=I(\rho_{AB})-\ CC(\rho_{AB}).$
If we denote $\mathrm{S_{\min} (\rho_{AB}) = \min_{\{B_k\}}
S(\rho_{AB}|\{B_k\}),}$ then a variant expression of $CC$ and $QD$ in Ref.\cite{J5,J12}
\begin{eqnarray}
CC(\rho_{AB})=S(\rho_{A})-min_{\{B_k\}}
S(\rho_{AB}|\{B_k\}),
\end{eqnarray}
\begin{eqnarray}
QD(\rho_{AB})=S(\rho_{B})-S(\rho_{AB})+S_{\min} (\rho_{AB}) .
\end{eqnarray}

\section{Effects of DM interaction on Various thermal correlations}
A typical solid-state system at thermal equilibrium in temperature $T$ (canonical ensemble) is $\rho(T)=e^{-\frac{H}{KT}}/Z$, with $Z=Tr[e^{-\frac{H}{KT}}]$ the partition function and $K$ is the Boltzmann constant. Usually we work with natural unit system $\hbar=K=1$ for simplicity and henceforth. This density matrix can be worked out as
\begin{equation}
\rho(T) =\frac{1}{Z} \left(
\begin{array}{cccc}
\rho_{11}&0&0&\rho_{41}\\
0&\rho_{22}&\rho_{23}^\dag&0\\
0&\rho_{23}&\rho_{22}&0\\
\rho_{41}&0&0&\rho_{11}
\end{array}
\right) ,
\end{equation}
where the elements of the matrix have been defined as
\begin{eqnarray*}
\label{eq}
\rho_{11}=\frac{1}{2}[e^{-\frac{J_x+J_y+J_z}{2T}}(e^{\frac{J_x}{T}}+e^{\frac{J_y}{T}})] ,
\end{eqnarray*}
\vspace*{-5mm}
\begin{eqnarray*}
\rho_{41}=\frac{1}{2}[e^{-\frac{J_x+J_y+J_z}{2T}}(-e^{\frac{J_x}{T}}+e^{\frac{J_y}{T}})] ,
\end{eqnarray*}
\vspace*{-5mm}
\begin{eqnarray*}
\rho_{22}=\frac{e^{{\frac{J_z}{2T}}}\cosh[\frac{\sqrt{(J_x +J_y)^{2}+4D^{2}_z}}{2T}]}{Z} ,
\end{eqnarray*}
\vspace*{-5mm}
\begin{eqnarray*}
\rho_{23}=\frac{(J_x+J_y-2iD_z)e^{{\frac{J_z}{2T}}}\sinh[\frac{\sqrt{(J_x +J_y)^{2}+4D^{2}_z}}{2T}]}{\sqrt{(J_x +J_y)^{2}+4D^{2}_z}Z},
\end{eqnarray*}
\vspace*{-5mm} and
\begin{eqnarray*} Z=e^{-\frac{J_x+J_y+J_z}{2T}}(e^{\frac{J_x}{T}}+e^{\frac{J_y}{T}})
\end{eqnarray*}
\vspace*{-5mm}
\begin{eqnarray*}
+2 e^{{\frac{J_z}{2T}}}\cosh[\frac{\sqrt{(J_x +J_y)^{2}+4D^{2}_z}}{2T}] .
\end{eqnarray*}

\begin{figure}[tbp]
\centering
\includegraphics[height=5.5cm,width=8cm]{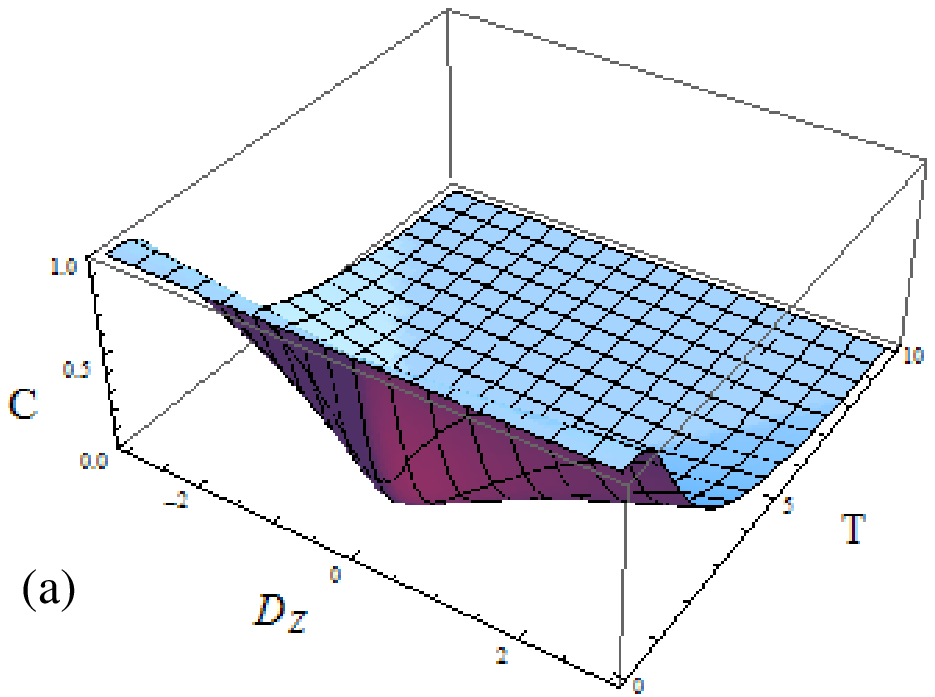}
\includegraphics[height=5.5cm,width=8cm]{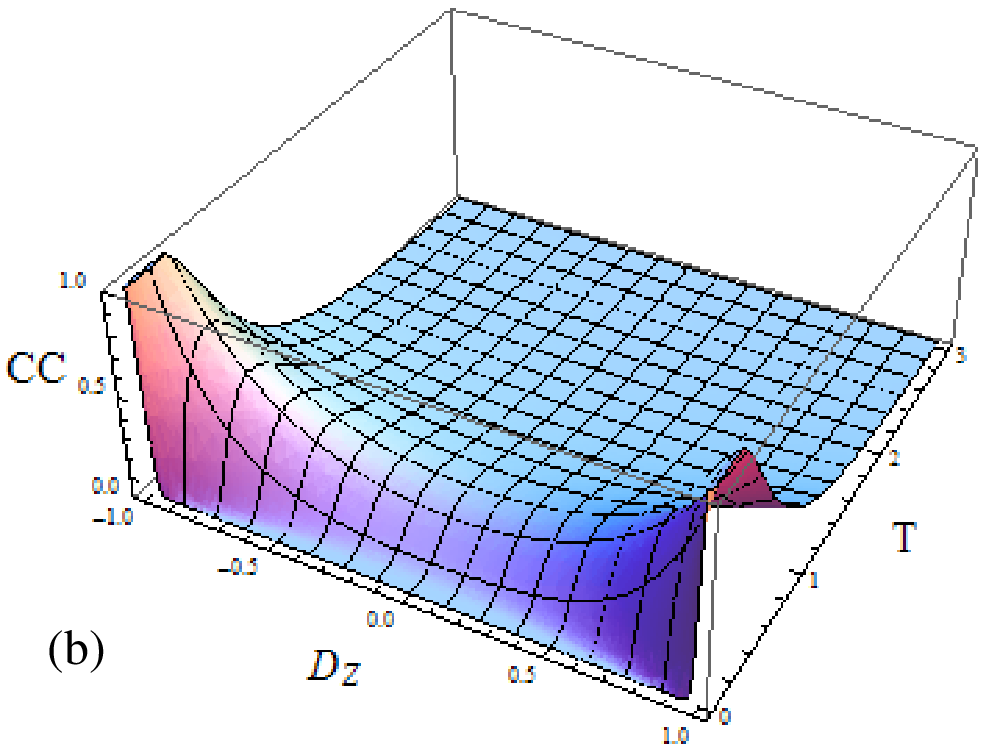}
\includegraphics[height=5.5cm,width=8cm]{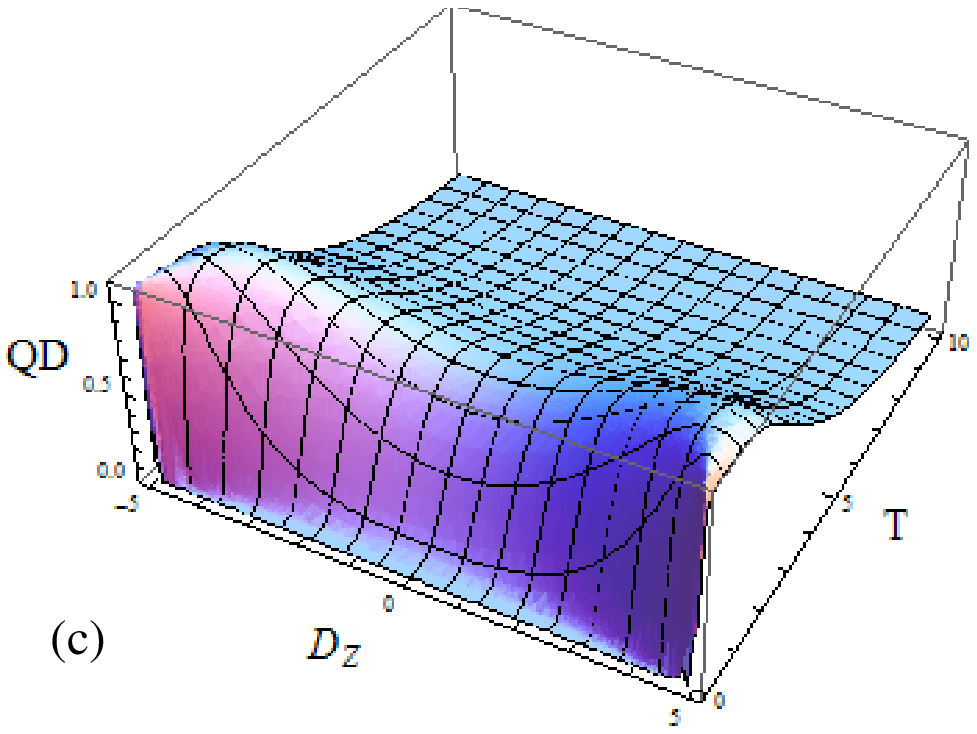}
\caption{(Color online) The concurrence (a), classical correlation (b) and quantum discord (c) versus $T$ and $D_z$. Here $J_x$=0.2, $J_y$=0.4, $J_z$=0.8.}\label{fig.1}
\end{figure}

According to the above definitions of $C$, $CC$ and $QD$, we will now discuss them with the corresponding plots. Fig. 1(a) shows that in the case of the temperature $T$ with finite value, $C(T)$ increases monotonously with the increasing of $D_z$, by which one can also achieve maximum entanglement even at finite low temperatures. Both $QD(T)$ and $CC(T)$ are zero when the temperature is zero, which is totally different from the case of $C(T)$ (it takes the maximum in this case). But there is an apparent increase, which is sharper for larger $D_z$, followed by a gradual decrease when the temperature is increased gradually starting from zero. More interestingly, the $QD(T)$ and $CC(T)$ show the same characteristics in their behavior following the increasing of the absolute value of $D_z$, which is different from the entanglement. The saddle-like structure of $QD(T)$ and $CC(T)$ in this case reveals the constructive role of $D_z$ for the two correlations, one quantum, one classical, which is one of the interesting results of this work. All of the above three correlations do not undergo sudden death, instead, they tend asymptotically towards zero as the temperature is increased.

In the overall, we conclude that $D_z$ is an efficient parameter in increasing various correlations such as $C$, $CC$ and $QD$ at finite temperature. This is partly contrary to the result for the case of XXZ model, in which the increase of the $DM$ interaction suppresses the $QD$ \cite{J38}. Moreover, the $QD$ shows a different behavior from the $C$ in the response to the variation of $DM$ interaction.

\section{Intrinsic Decoherence of Various Correlations}
 Now, we take the influence of intrinsic decoherence on the various correlations into account. According to the Milburn's equation \cite{J39} followed by the assumption that a system does not evolve continuously under unitary transformation for sufficiently short time steps, the master equation for pure phase decoherence is given by
\begin{eqnarray}
\frac{d\rho(t)}{dt} = -i[H, \rho] - \frac{1}{2\gamma}
[H,[H,\rho(t)]],
\end{eqnarray}
where $\gamma$ is the phase decoherence rate. In the limit $\gamma\rightarrow 0$ the Schodinger$^{,}$s equation is recovered.
The formal solution of the master equation above can be given by \cite{J40}
\begin{eqnarray}
\rho(t) = \sum_{k=0}^{\infty} \frac{l^k}{k!} M^{k}(t) \rho(0) M^{\dagger k}(t),
\end{eqnarray}
where $\rho(0)$ is the density operator of the initial system and
$M^{k}(t)$ is defined by
\begin{eqnarray}
M^{k}(t)= H^{k} e^{-iH t} e^{-\frac{t}{2\gamma} H^2
}.
\end{eqnarray}
By inserting the completeness relation $\sum _{n}| \psi_n \rangle \langle \psi_n |=1$ of the energy eigenstate into master equation \cite{J39}, we can write the explicit expression of the density matrix of the states as
\begin{eqnarray*}
\hspace*{6mm}
\rho(t) =\sum_{mn}\mathrm{exp} \bigg[-\frac{\gamma t}{2} (E_m - E_n)^2 -i(E_m - E_n) t
\bigg] \qquad
\end{eqnarray*}
\vspace*{-2mm}
\begin{eqnarray}
 \qquad\hspace*{-18mm}\times \langle \psi_m | \rho(0) | \psi_n \rangle | \psi_m \rangle \langle \psi_n| .
\end{eqnarray}
We assume that the system is initially prepared in the Bell state $|\Psi(0)\rangle =\frac{1}{\sqrt{2}} \big(| 01 \rangle + | 10
\rangle \big)$. From the Eq. (16), the time evolution for this initial state can be obtained as
\begin{equation}
\rho(t) =\left(
\begin{array}{cccc}
0&0&0&0\\
0&\rho_{22}&\rho_{23}&0\\
0&\rho_{32}&\rho_{33}&0\\
0&0&0&0
\end{array}
\right) ,
\end{equation}
where the elements of the matrix can be defined as
\begin{eqnarray*}
\label{eq}
\rho_{22}=\frac{1}{2}+\frac{D_z e^{-\frac{1}{2}\gamma\mu^2 t}\sin[\mu t]}{\mu^2},
\end{eqnarray*}
\vspace*{-5mm}
\begin{eqnarray*}
\rho_{23}=\frac{J_x+J_y-2iD_z e^{-\frac{1}{2}t\gamma\mu^2}\cos[t\mu]}{2(J_x+J_y-2iD_z)},
\end{eqnarray*}
\vspace*{-5mm}
\begin{eqnarray*}
\rho_{32}=\frac{J_x+J_y+2iD_z e^{-\frac{1}{2}\gamma\mu^2 t}\cos[\mu t]}{2(J_x+J_y+2iD_z)},
\end{eqnarray*}
\vspace*{-5mm}
\begin{eqnarray*}
\rho_{33}=\frac{1}{2}-\frac{D_z e^{-\frac{1}{2}\gamma\mu^2 t}\sin[\mu t]}{\mu^2}
\end{eqnarray*}
\vspace*{-5mm} and where
\begin{eqnarray*}
\mu=\sqrt{J^2_x+2J_x J_y+J^2_y+4D^2_z} .
\end{eqnarray*}

\begin{figure}[tbp]
\centering
\includegraphics[height=6cm,width=8cm]{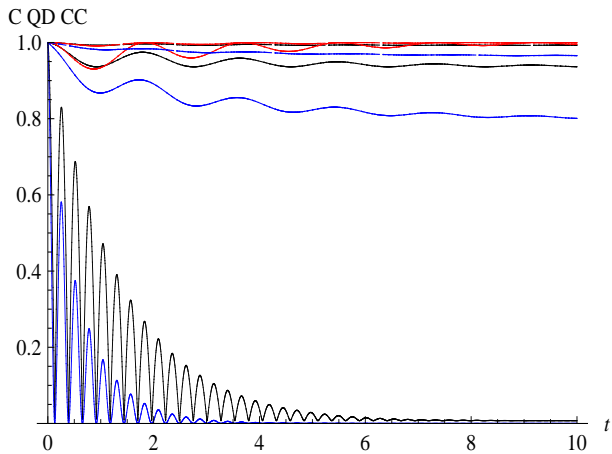}
\caption{(Color online) The lower part of the figure is the concurrence (black line), quantum discord (blue line) versus time $t$ with the system parameters fixed as $J_x$=0.03, $J_y$=0.06, $\gamma=0.01$ and $D_z$=6. The upper part of the figure is the concurrence (black line), quantum discord (blue line), classical correlation (red line) versus time $t$ with the system parameters fixed as $J_x$=3, $J_y$=0.6, $\gamma=0.1$ and $D_z$=0.1 (dotted line), $D_z$=0.3 (thin line).}\label{fig.2}
\end{figure}

In order to highlight the effect of the pase decoherence $\gamma$ on the various correlations, we plot the time evolutions of correlations with different values of $D_z$ in Fig.\,2. It can be seen from the lower part of the figure that the time evolution of the entanglement and quantum discord exhibit the interesting phenomena of "sudden death" and "sudden revival",\cite{J4} which occur when the spin-orbit coupling is large and spin-spin coupling is small. Secondly, with the aim to clarify the joint influence of the system parameters with the phase decoherence on the time evolution, the combination of the system parameters for the upper part of the figure is chosen as the optimum one based on the numerical analysis. All of the three correlations exhibit oscillatory behavior, which ultimately ends with a steady state value. Oscillations are suppressed obviously with the increase of $\gamma$. The CC ends with the maximum value, while the other two end with a smaller steady state value with respect to the starting maximum value.
Importantly, the final steady state values of the entanglement and quantum discord are all still high despite the phase decoherence, implying that the optimum combination of the system parameters can keep the correlations highly immune to the pure phase decoherence. Moreover, one can see that the quantum discord is more fragile under the phase decoherence than the entanglement, which is different from the result that it is more resistant against the environment than entanglement. Last but not least, the larger the DM interaction, the severer the collapse of the correlations, which is the opposite of the thermal case.

\section{Conclusion}
In conclusion, we have studied the various correlations, particularly the quantum discord in an anisotropic two-qubit Heisenberg XYZ model with the presence of DM interaction. Results are presented both for the case at thermal equilibrium and under phase decoherence and they show that the roles of the DM interactions in controlling the thermal quantum discord are opposite to the case of the XXZ.
It is found constructive in the case of XYZ model under our consideration. However, this is not the same story for the case of phase decoherence, where the DM becomes destructive. The time evolution of the entanglement and quantum discord shows the famous phenomena of collapse and revival. Though the quantum discord is shown to be more sensitive to the phase decoherence than the entanglement, optimum combination of the system parameters can protect the correlations effectively against the influence of the phase decoherence on the whole.

\end{document}